\documentclass[aps,twocolumn,epsfig,graphics,showpacs,floatfix,mathbbm]{revtex4}

\usepackage{amsmath,amsfonts,amssymb,graphics,graphicx,epsfig,color,times,bbm}

\begin{document}

\bibliographystyle{apsrev}

\newtheorem{theorem}{Theorem}
\newtheorem{proposition}{Proposition}
\newtheorem{lemma}{Lemma}
\newcommand{\proofend}{\hfill\fbox\\\medskip }
\newcommand{\proof}[1]{{\bf Proof.} #1 $\proofend$}
\newcommand{\rr}{\mathbbm{R}}
\newcommand{\nn}{\mathbbm{N}}
\newcommand{\cc}{\mathbbm{C}}
\newcommand{\qq}{\mathbbm{Q}}
\newcommand{\id}{\mathbbm{1}}
\newcommand{\tr}{\text{tr}}
\newcommand{\Bra}[1]{\ensuremath{\langle#1|}}
\newcommand{\bra}[1]{\ensuremath{\langle#1|}}
\newcommand{\Ket}[1]{\ensuremath{|#1\rangle}}
\newcommand{\ket}[1]{\ensuremath{|#1\rangle}}
\newcommand{\BraKet}[2]{\ensuremath{\langle #1|#2\rangle}}
\newcommand{\braket}[2]{\ensuremath{\langle #1|#2\rangle}}
\newcommand{\KetBra}[1]{\ensuremath{| #1 \rangle \langle #1 |}}
\newcommand{\ketbra}[1]{\ensuremath{| #1 \rangle \langle #1 |}}
\newcommand{\KetBraO}[3]{\ensuremath{| #1 \rangle_{#3}\langle #2 |}}
\newcommand{\Eins}{\ensuremath{\mathbbm 1}}
\newcommand{\eins}{\ensuremath{\mathbbm 1}}
\newcommand{\WW}{\ensuremath{\mathcal{W}}}
\newcommand{\BE}{\begin{equation}}
\newcommand{\EE}{\end{equation}}
\newcommand{\be}{\begin{equation}}
\newcommand{\ee}{\end{equation}}
\newcommand{\bea}{\begin{eqnarray}}
\newcommand{\eea}{\end{eqnarray}}

\title{Complete hierarchies of
efficient approximations to 
problems in entanglement theory}

\author{Jens Eisert$^{1,2}$, Philipp Hyllus$^3$, Otfried G{\"u}hne$^{3,4}$, and
Marcos  Curty$^5$ }

\affiliation{
1 
Institut f{\"u}r Physik, Universit{\"a}t Potsdam,
Am Neuen Palais 10, 14469 Potsdam, Germany\\
2 Blackett Laboratory, Imperial College London,
Prince Consort Road, London SW7 2BW, UK\\
3 Institut f{\"u}r Theoretische Physik,  Universit{\"a}t  Hannover,
  Appelstra{\ss}e 2,  30167 Hannover,  Germany\\
4 Institut f{\"u}r Quantenoptik und Quanteninformation,
{\"O}sterreichische Akademie der Wissenschaften,
     6020 Innsbruck,     Austria   \\
5 Institut f{\"u}r Theoretische Physik I and Max-Planck 
Research Group,
Institute of Optics, Information and Photonics, Universit{\"a}t 
Erlangen-N{\"u}rnberg,
Staudtstra{\ss}e 7/B2, 
91058 Erlangen, 
Germany                                        
}
\date{\today}

\begin{abstract}
We investigate several problems in entanglement theory
from the perspective of convex optimization. This list of problems 
comprises (A)  the decision whether a state is  multi-party entangled, 
(B) the minimization of expectation values 
 of entanglement witnesses with respect to pure product states, 
(C) the closely related evaluation of the 
geometric measure of entanglement to quantify pure 
multi-party entanglement, (D) 
the test whether states are multi-party entangled on the basis 
of witnesses based on second moments and on the basis of 
linear entropic criteria, and (E) the evaluation of instances of 
maximal output purities of quantum channels.
We show that these problems can be formulated as certain 
optimization problems: as polynomially
constrained problems employing polynomials of degree three or less. 
We then apply very recently established known
methods from the theory of semi-definite relaxations to the formulated 
optimization problems. By this construction we arrive at a
hierarchy of efficiently solvable approximations to the solution, 
approximating the exact solution as closely as desired, 
in a way that is asymptotically complete. 
For example, this results in a hierarchy of novel, 
efficiently decidable sufficient criteria for multi-particle entanglement, 
such that every entangled state will necessarily be detected in some 
step of the hierarchy. Finally, we present numerical examples to 
demonstrate the practical accessibility of this approach.
\end{abstract}

\maketitle

\section{Introduction}

One of the reasons for the superior performance of quantum devices
for computation and communication 
compared to their classical counterparts is simply due to the fact that 
in quantum mechanics, one has a very large space at hand to work with:
the dimension of the state space of a number of quantum bits 
is expontially larger than the corresponding configuration space
of classical bits. This renders the simulation of a quantum computer on a 
classical device a difficult task. 
But it is not only the sheer size of state space 
that makes the assessment of quantum states a difficult problem.
In fact, even 
to decide whether quantum states
have certain properties that are of central interest in quantum 
information science often 
amounts to solving computationally hard problems 
on a classical computer. 
Most prominently, 
to decide whether a known state $\rho$
of a finite-dimensional 
bi-partite system is separable or entangled, i.e., 
whether or not it can be written as a convex combination of 
product states
\begin{equation}\label{cc}
	\rho= \sum_{i=1}^n p_i \rho_1^{(i)}\otimes  \rho_2^{(i)},
\end{equation}
is already an NP-hard problem in the system size 
\cite{Gurvits}. 
A state is separable
if there is a preparation of the state that involves only local
quantum operations and shared classical randomness.  Such states
are correlated, but classically correlated, as the source for the 
correlations
can be thought of as resulting entirely 
from the shared source of 
randomness  \cite{Wer89}. 
Due to the central status of the concept of entanglement in quantum 
information, a very significant amount of research has been dedicated to 
the
problem of finding good criteria for separability that are suitable for 
specific contexts \cite{Separability}. 

To state whether a state is separable or not is equivalent to 
stating whether
a state is in the convex hull of product states. Also, the evaluation of 
many measures of entanglement
essentially require the solution of a convex problem. So in 
recent years, it has increasingly been realized that a good deal of
new insight in several problems in quantum information 
and in particular in entanglement theory could in fact 
come from the field of research that is primarily concerned
with questions of this type 
\cite{ConvexButNotSemidef,Rains,ConvexInQuantumInfo,Global,Doherty,Brasilians,Gurvits,Ioannou}: 
this is the theory 
of convex optimization. Many problems are
already of the required form, and powerful tools such as the concept of
Lagrange duality readily deliver bounds for the problems at hand.  Examples
include the evaluation of measures of entanglement that reasonably quantify
the degree of entanglement of a given state, such as the distillable 
entanglement or the asymptotic relative entropy of entanglement
\cite{Rains,ConvexButNotSemidef}. 
Also, it has been realized that while the complete solution of the 
question of separability is NP-hard, one can nevertheless
find hierarchies of sufficient criteria for entanglement in the bi-partite
setting. In each step, by
solving an efficiently solvable convex optimization problem, one finds
an answer to the problem in the form (i) one can assert that the state
is entangled, or (ii) one cannot assert it, and has to go one 
(computationally
more expensive) step further \cite{Doherty}. The problem of testing 
for multi-partite entanglement 
has been related to robust 
semi-definite programming and a hierarchy of relaxations
in Ref.\ \cite{Brasilians}. 

This paper is concerned with a link of the theory of entanglement to
the theory of convex optimization in a similar
spirit. The central observation of this paper is very
simple yet potentially very useful:
many problems related to entanglement
can be cast into the form of optimization problems with
polynomial constraints of degree three.
This includes the (A) 
question whether
a state is entangled or not, notably not only in the bi-partite, but
also for the several separability classes of the multi-partite setting.
Then, (B) the construction of non-decomposable witnesses involves
a problem of this kind, as well as the (C) evaluation of the geometric
measure of entanglement to quantify multi-partite entanglement. (D) Also, 
when considering entanglement witnesses based on second moments rather 
than on first moments one has to solve a problem of this form. We will 
also discuss
criteria based on linear entropies (i.e., $p$-norms for $p=2$). (E)
Finally, we will briefly mention the evaluation of maximal output purities 
of quantum channels
with respect to $p$-norms for $p=2$.
This structure is due to the fact that in all these instances, one 
essentially
minimizes over product state vectors of a multi-partite quantum system.

This polynomial part  of the optimization problems is 
still non-convex and computationally expensive to solve. 
Yet, applying  
results from relaxation theory of non-convex problems
\cite{Shor,Kojima,Lasserre,Lasserre2,Par}, notably the method of Lasserre 
\cite{Lasserre},
we find hierarchies of 
solutions to our original problems, and each step is a better 
approximation 
than the previous one. Each step itself amounts to solving an efficiently
implementable semi-definite program \cite{Semi}. Moreover, the hierarchy is 
asymptotically
complete, in the sense that the exact solution is asympotically attained. 
The increase of the size of the vector of objective variables of these
semi-definite problems grows notably polynomially in the label of the 
hierarchy.

We will first clearly state how one can introduce auxiliary variables to 
cast the considered problems from entanglement theory 
into the desired form. Then, we will investigate 
the hierarchies of relaxations in 
detail, and study numerical examples. 
Finally, we will summize what has 
been achieved. 

\section{Problems in entanglement theory 
as optimization problems}\label{sec:2} 

The problems that we will encounter are of the following type
or similar. At the core are typically minimizations over product vectors,
originating  from the very definition of the concept of entanglement. 
Given a  $W=W^\dagger$, we seek the minimum of 
\begin{equation}\label{optigleichung}
	\tr[|\psi_1\rangle \langle \psi_1| \otimes ...\otimes
	|\psi_N\rangle\langle
	 \psi_N| W],
\end{equation}
where the minimum is taken with respect to product state
vectors of a composite quantum systems with
parts labeled $1,...,N$, with Hilbert space ${\cal H}={\cal H}_1\otimes ...
 \otimes {\cal H}_N$. Throughout the paper, 
the respective Hilbert spaces
are assumed to have finite dimensions, ${\cal H}_j=
\cc^{d_j}$, $j=1,...,N$.

One way
of solving this problem is to choose a specific basis for 
 the Hilbert space and to
explicitly parametrize
the state vectors. This yields a complex
polynomial in these parameters, in general of 
very high order. This is obviously 
not a convex problem in these
variables: a solution can be found, albeit not in an efficient
manner. For small systems, algorithms such as simulated
annealing may be employed, 
delivering upper bounds to the optimal 
solution, as no control is possible as to what extent one is far
away from the global optimum. 

The general strategy of this paper is in instances
of the above type to introduce additional variables, 
giving rise to one vector $x\in \rr^t$, $x=(x_1,...,x_t)^T$, 
which is the objective
variable, parametrizing the product states. 
The problem is  then cast into 
the form of a linear objective function, simply as
\begin{equation}
	\text{minimize } c^T x
\end{equation}
with a (fixed) $c\in \rr^t$, 
subject to constraints which are
polynomials in the objective variables. These constraints
will then be relaxed to semi-definite problems. So two types of
contraints will be encountered in the present paper:
\begin{itemize}
\item  {\it Semi-definite 
constraints:} 
These are constraints of the form
\begin{equation}
 	F_0 + \sum_{s=1}^t 
	x_s F_s\geq 0
\end{equation}
where $F_0,...,F_t$ are Hermitean matrices of arbitrary dimensions.
The resulting matrix has to be positive semi-definite, therefore, it is 
referred to
as semi-definite
constraint. Optimization problems of this type, 
exhibiting a linear objective function and semi-definite
constraints, are called semi-definite programs \cite{Semi}. 
Such instances of convex 
optimization 
problems can be efficiently solved, for example by means of interior-point 
methods \cite{Semi}. Moreover, the idea of Lagrange duality 
 \cite{LagrangeDuality}
 readily 
delivers
lower bounds for the problem. 
Typically, the dual optimization problem 
yields
an optimal value which is identical to the optimal value for the primal
problem
(unless there is a duality gap). Many problems in quantum information 
theory
have already the form of a semi-definite program 
\cite{ConvexInQuantumInfo,Doherty}. 
In fact, it may be convincingly argued that 
to specify the
solution of a problem in form of a semi-definite program has the same 
status as stating a result in terms of the spectrum of a matrix, as this 
again merely means that efficient methods are available to find the
eigenvalues of a given matrix.

\item {\it Polynomial constraints:}
This means that we can write the constraints as
\begin{equation}\label{Quad}
	g_l(x)\leq 0,
\end{equation}
$l=1,...,L$, where $g_l:\rr^t\rightarrow \rr$ are real polynomials of some 
degree. Quadratic constraints are
of the form 
\begin{equation}\label{Quad}
	x^T A_l x + b_l^T x + c_l\leq 0,
\end{equation}
$l=1,...,L$. The matrices $A_l$ are, however, not necessarily 
positive semi-definite. This is by no means a minor detail: 
if all matrices $A_1,...,A_L$ were positive matrices, $A_l\geq 0$,
this would
yield a convex quadratic program, 
which can be efficiently
solved (they are in fact also instances of semi-definite programs
and of second-order cone programs).
In stark contrast, if the matrices are not all 
positive semi-definite, one obtains
a very hard, non-convex optimization problem. This structure
is yet dictated by the problems from quantum information theory
at hand. 
\end{itemize}

The central point is to employ known methods from 
the theory of relaxations of non-convex optimization problems, 
to obtain complete hierarchies of cheaply computable
approximations, 
approximating the solution as closely as desired. 
The idea of a relaxation is to
introduce new variables and to formulate the problem as a convex problem in a 
larger space. This idea can be exemplified in the simplest form of a 
relaxation,
the Shor relaxation \cite{Shor}. 
For example, let $A_1$ in Ineq.\ (\ref{Quad})
be a matrix which is not positive 
semi-definite, and let us assume that 
$b_1=0$ and $c_1=0$ for simplicity. 
Then, one can still write the
constraint equivalently
as
\begin{equation}
	\tr [X A_1] \leq 0,\,\, X=x x^T,
\end{equation}
using a $t\times t$ symmetric matrix $X$. The equality
$X=x x^T$ is equivalent with the convex constraint
\begin{equation}
X\geq x x^T,
\end{equation} 
together with 
the non-convex one $X\leq x x^T$. Shor's relaxation amounts
to taking only the convex part into account, thereby delivering an
efficiently solvable convex problem which yields a lower bound to
the original problem \cite{Shor}. 
Such relaxations in terms of semi-definite constraints
will be employed, yet instead of one 
many such relaxations forming a complete hierarchy.

As pointed out before, we will show that the encountered optimization
problems can be written as polynomially constrained
problems of degree three. 
That this is possible is based on the observation that 
any Hermitean 
$m\times m$-matrix  $O$ for which 
\begin{equation}
	\tr[O^2]=1, \, \tr[O^3]=1
\end{equation}
is one that satisfies
\begin{equation}
	\tr[O]=1, \, O=O^2, \, O\geq 0,
\end{equation}
i.e., it corresponds to a pure state (see also Ref.\ \cite{NewJones}). 
This follows from the fact that, 
denoting the
decreasingly 
ordered list of eigenvalues of $O$ with $\lambda^\downarrow(O)$, the only
vector consistent with
\begin{eqnarray}
	\sum_{i=1}^m 
	\lambda^\downarrow_i(O)^2 =1, \,\,\,\sum_{i=1}^m 
	\lambda^\downarrow_i(O)^3=1 
\end{eqnarray}
is the vector $\lambda^\downarrow(O)=(1,0,...,0)$. The  quantities
$\lambda^\downarrow_i(O)^2 $ and $\lambda^\downarrow_i(O)^3$
are unitarily invariant, and hence, the above 
statements can be shown to be valid
on the level of probability distributions.  Essentially, $\sum_{i=1}^m \lambda^\downarrow_i(O)^2=1$
already requires all absolute values of eigenvalues to be smaller than
or equal to $1$, such that
the only ordered vector of real numbers consistent with $\sum_{i=1}^m \lambda^\downarrow_i(O)^3=1$
becomes $(1,0,...,0)$.

For systems where the individual constituents are qubit systems, $d_j=2$ 
for all $j=1,...,N$,
the constraints  can further be simplified by merely requiring as 
constraints
$\tr[O]=1$, $\tr[O^2]=1$,
as  for Hermitean $2\times 2$-matrices these 
conditions
alone
imply that  
\begin{equation}
O\geq 0,\,\,  O=O^2.
\end{equation}

When applied to our specific problems at hand, these constraints 
will appear in the following form. We will require that Hermitean 
matrices $P$ are, except from normalization, products of pure states
with respect to all constituents. This will be incorporated as follows: 
Denoting with
$I=\{1,...,N\}$ the index set labeling the subsystems
and with $ \tr_{I\backslash j}$ the partial trace 
with respect to all systems except the one with label 
$j$,
the lines 
\begin{eqnarray}
\tr[ \tr_{I\backslash j} [P ]^2 ] &=&  (\tr [P  ])^2,\\
\tr[ \tr_{I\backslash j} [P ]^3 ] &=&  (\tr [P  ])^3
\end{eqnarray}
for all $j\in I$ 
indeed enforce that
the matrices are products. If reductions are pure, the global
state must be a pure product state. 
This can be seen as follows. For states $\rho$, 
the only possibility for 
\begin{eqnarray}
	\tr[ \tr_{I\backslash j} [\rho]^2 ]  = 1,\,\,\,
	\tr[ \tr_{I\backslash j} [\rho]^3 ]  = 1
\end{eqnarray}
to hold for all $j\in I$ is that $\rho$ is of the form
of a product pure 
state,
\begin{equation}
\rho=|\phi_1\rangle\langle\phi_1|\otimes ...\otimes 
|\phi_N\rangle\langle\phi_N|.
\end{equation} 
If an additional constant 
$\alpha>0$ is included, 
these conditions read 
$\tr[ \tr_{I\backslash j} [\alpha\rho]^2 ]=  
(\tr 
[\alpha\rho])^2= \alpha^2$ and 
$\tr[ \tr_{I\backslash j} [\alpha\rho]^3 ]=   (\tr 
[\alpha\rho])^3= \alpha^3$, 
which explains the above constraint  \cite{QubitRemark}.
Having stated the general strategy, 
let us now look at the specific
instances of problems in quantum information we will be considering in this
paper. 

\subsection{Tests for bi-partite and multi-partite entanglement}

The approach 
is here to consider for a given state
$\rho\in {\cal S}({\cal H}_1\otimes ...\otimes {\cal H}_N)$ 
the minimal Hilbert-Schmidt norm with
respect to the set of separable states. 
For simplicity of notation, we explicitly
formulate the optimization problem for the instance of full separability, 
without loss
of generality. 
That is, we
test whether $\rho$ can be written as
\begin{equation}\label{cc}
	\rho= \sum_{i=1}^n p_i \rho_1^{(i)}\otimes ...\otimes  \rho_N^{(i)},
\end{equation}
with $\{p_i\}_i$ forming a probability distribution. 
The question whether a state is fully separable is hence 
equivalent to
asking whether a state is an element of the convex hull of 
product vectors with respect to all subsystems.
According to Caratheodory's theorem \cite{Rocky}, 
for any $k$-dimensional subset $S\subset\rr^m$, any point of the convex 
hull
of $S$ can be written as a convex combination of at most $k+1$
points from $S$. Hence, the 
number of elements in the convex combination 
given by Eq.\ (\ref{cc}) can be 
restricted to $n=\prod_{j=1}^N d_j^2$, again 
without loss of generality. 
To decide whether a state $\rho$ is fully separable or not, we may
solve the following optimization problem,
\begin{eqnarray}\label{themin}
	\text{minimize}&& \| \rho - P \|_2^2= \tr (\rho-P)^2, \\
	\text{subject to} && P \text{ is fully separable}. \nonumber
\end{eqnarray}
We make use of the Hilbert-Schmidt norm as it is quadratic
in the matrix entries.

The task is  to write this problem in terms of 
a polynomially constrained problem. 
Each relaxation (see Section \ref{boyrelax}), 
labeled with $h = h_{\text{min}}, h_{\text{min}}+1,...$, 
then delivers a 
lower bound of the Hilbert Schmidt distance to the set of fully separable
states. Hence, asserting that the state is not fully separable whenever
we obtain a value larger than the one that we accept as accuracy of 
the computation \cite{Sharp}, each step delivers a
sufficient criterion for multi-partite entanglement in its own right, and
the hierarchy is complete in the sense that each entangled state
is detected by some step. 
The  associated optimization problem can now be written as
\begin{eqnarray}\label{withsemi}
	\text{minimize}&& x,\\
	\text{subject to} && 
	x\geq\tr (\rho  -   P)^2 ,\nonumber\\
	&&P-\sum_{i=1}^n P^{(i)} =0,\nonumber\\
	&&\tr[ \tr_{I\backslash j} [P^{(i)}]^2 ] =  (\tr [P^{(i)} ])^2 
\,\nonumber\\
		&&\hspace*{1cm}
		\text{ for all } i=1,...,n,\, j\in I,\nonumber\\
	&&\sum_{i=1}^n \tr[P^{(i)}]=1,  \nonumber\\
	&& P^{(i)}\geq 0,\, \text{ for all }  i=1,...,n.  \nonumber
\end{eqnarray}
The line $\sum_{i=1}^n \tr[P^{(i)}]=1$ 
takes the normalization of the
whole state into account.
This is a quadratic program, combined with a semi-definite
constraint for the positivity of the matrices $P^{(i)}$. 
As pointed out above, the problem can also
be formulated as a polynomial problem without a semi-definite
constraint, but now with constraints that are of degree three. 
\begin{eqnarray}
	\text{minimize}&& x,\\
	\text{subject to} && 
	x\geq\tr (\rho  -   P)^2 ,\nonumber\\
	&&P-\sum_{i=1}^n P^{(i)} =0,\nonumber\\
	&&\tr[ \tr_{I\backslash j} [P^{(i)}]^2 ] =  (\tr [P^{(i)} ])^2,  
\,\nonumber\\
	&&\hspace*{1cm}
		\text{ for all } i=1,...,n,\, j\in I . \nonumber\\
		&& \tr[ \tr_{I\backslash j} [P^{(i)}]^3 ] =  (\tr 
[P^{(i)} ])^3,  \,\nonumber\\
		&&\hspace*{1cm}
		\text{ for all } i=1,...,n,\, j\in I . \nonumber\\
	&& \sum_{i=1}^n \tr[P^{(i)}]=1.  \nonumber
\end{eqnarray}
This is a global optimization problem with polynomial constraints of degree
three, but no semi-definite constraint.

Here one tests the hypothesis that the state is fully 
separable against the alternative that the state is 
entangled in some sense. To assert that the state
is multi-particle entangled and not separable with respect to any 
separability class, several tests are hence required. In this way,
the various classes of genuine multi-particle entanglement can be 
detected. Note that when even applied to the bi-partite case, the
resulting hierarchy of semi-definite relaxations is inequivalent to
the one in Ref.\ \cite{Doherty}, and also inequivalent to the
robust semi-definite programming approach in Refs.\ \cite{Brasilians}.
The above  
formulation in the optimization problem in terms of full separability
still does not constitute a restriction of generality, as this includes 
all separability classes with respect to all possible splits. 

Alternatively to the above approach, 
one may write each test in the form of a feasibility 
problem, a problem with a vanishing objective function,
\begin{eqnarray}\label{themin}
	\text{minimize}&& 0, \\
	\text{subject to} && \rho \,\, \text{ satisfies the test of 
step}\nonumber\\
	&& \hspace*{.35cm} \text{$h=h_{\text{min}},h_{\text{min}+1},... $ in the hierarchy}. \nonumber
\end{eqnarray}
Either one finds no solution (which is to say, 
the problem is not primal feasible),
and one can assert that the state is not
fully separable, or one has to go on one step in the hierarchy. 
In each step of the hierarchy, forming a
semi-definite problem, the dual 
problem can then be employed to prove the infeasibility of the above primal 
problem serving as a certificate \cite{Semi} (see also Ref.\ 
\cite{Doherty}).  

In general the total 
problem can in each step  be written as a semi-definite
problem of the form
\begin{eqnarray}
\text{minimize} && 0,\\
\text{subject to} && H_0 + \sum_{s=1}^T z_s H_s\geq 0,\nonumber
\end{eqnarray}
with appropriate matrices $H_s$, $s=1,...,T$.
The associated Lagrange 
dual problem  \cite{LagrangeDuality}
is again a semi-definite program, 
\begin{eqnarray}
\text{maximize} && -\tr[Z H_0],\\
\text{subject to} && \tr[Z H_s]=0,\, s=1,...,T,\nonumber\\
&& Z\geq 0.\nonumber
\end{eqnarray}
In the context of our feasibility problem above, 
any feasible solution of the dual problem 
with $ \tr[Z H_0]<0$ proves the infeasibility of the primal (original)
problem. That is, we can use the dual problem to prove properties of
our original problem at hand.

Finally, it is important to point out that in problem (\ref{withsemi}),
one may keep the semi-definite constraint provided by the last line,
and look for the intersection
of the feasible sets of the semi-definite part and the constraint set of 
the relaxations. Then, in each step we can either assert whether the 
state is entangled, or one cannot say whether it is entangled or not.
In this way, it may happen, yet, that the state is entangled, although this
entanglement is not detected in any step of the hierarchy. 
One hence obtains a hierarchy of sufficient criteria, albeit one
which is not  
necessarily asymptotically complete. 
This will be discussed in more detail in the
section on the hierarchy of relaxations.

In an implementation of this optimization problem, one
has to choose a basis of 
Hermitean matrices for each Hilbert space, 
\begin{equation}
\{ \sigma_1,...,\sigma_{d_j^2}\},
\end{equation} 
for $j=1,...,N$, suppressing an additional index labelling the subsystems. 
These
Hermitean matrices
satisfy $\tr[\sigma_1]=1$ and
\begin{equation}
	  \tr[\sigma_k]=0,\,\,k=2,...,d_j^2, 
\end{equation}	
	and have a Hilbert-Schmidt scalar product
\begin{equation}
	\tr[\sigma_k \sigma_l ]= \xi_{d_j} \delta_{kl}
\end{equation} 
with a dimension dependent
constant $\xi_{d_j}$ (and similarly for terms of third order). 
For the case of qubit subsystems, the appropriately
normalized familiar Pauli matrices can be taken. 
In terms of this basis of Hermitean matrices, 
the matrices $P^{(i)}$ and $P$
can be written as
\begin{eqnarray}
P^{(i)}&=& \sum_{\kappa=(k_1,...,k_N)} p_{\kappa}^{(i)} \Sigma_\kappa,\\
P&=& \sum_{\kappa=(k_1,...,k_N)} p_{\kappa} \Sigma_\kappa,
\end{eqnarray}
where $\kappa=(k_1,...,k_N)$,   is a multi-index, with $k_j=1,...,d_j^2$ 
for $j\in I$, and 
\begin{equation}
	\Sigma_\kappa= \sigma_{k_1}\otimes \sigma_{k_2}\otimes ...\otimes
\sigma_{k_N}.
\end{equation}
This parametrization will be used in the section
presenting numerical examples. Before we present
the hierarchy of relaxations explicitly, 
we discuss the other applications which are
similar in structure from the point of view taken in this paper.

\subsection{Non-decomposable witnesses}
Optimization problems of the type of the one in Eq.\ (\ref{optigleichung})
often appear in the construction of entanglement 
witnesses \cite{witness}. An entanglement 
witness is a Hermitean observable $W=W^\dagger$ with the property 
that $\tr[W\rho]\geq 0$ holds for all separable $\rho$, 
thus a negative expectation value signals the presence of 
entanglement. 
So entanglement witnesses can be used for an   experimental  
verification that a given state is entangled, and, in fact, 
they have already been implemented \cite{barbieri}.

The detection of entanglement is not only of interest 
for fundamental reasons, it can also be of practical 
interest. This is the case in quantum crytography, since 
it has been shown that the provable presence of quantum 
correlations in such protocols is a necessary precondition 
for secure key distillation \cite{curtyPRL}. 
Entanglement witnesses are particularly suited to deliver this 
entanglement proof, even when the quantum state shared by the users 
cannot be completely reconstructed. In turn, 
by measuring {\it all} accessible 
witnesses, one can decide whether the measurable correlations of the
state origin from an entangled state or may be compatible with a 
separable state. 

There are many strategies to construct entanglement 
witnesses \cite{optimization,constructW}. As an example in which 
such optimisation problems occur we choose the construction 
of non-decomposable witnesses for PPT entangled states 
\cite{optimization}. These are entangled states which have a
positive partial transpose \cite{PPTRemark}.
We discuss our example in the bi-partite setting for simplicity.
In the theory of PPT entangled states the extreme points of the set
are of central interest, then often referred to as
edge states. A state $\rho$ is a PPT entangled edge state
if it has a positive partial transpose, while for all product 
vectors $\ket{a,b}$  in the range of $\rho$ 
the vector $\ket{a,b^*}$ is not in the range 
of the partially transposed $\rho^{T_B}$ \cite{witness,optimization}.
Here $*$ refers to complex conjugation. 

To construct witnesses for these states, one proceeds as 
follows: 
Let $R=K(\rho)$ and $Q=K(\rho^{T_B})$ be the projectors onto 
the kernels of $\rho$ and of $\rho^{T_B}$, respectively.
Then a witness allowing the detection of the state $\rho$ is given 
by \cite{optimization}
\begin{equation}
W' = R+Q^{T_B}- \varepsilon \eins
\end{equation}
where
\begin{eqnarray}
\varepsilon
&=&\min_{\ket{a,b}} \tr[\ketbra{a,b}(R+Q^{T_B})].
\label{witnessmin}
\end{eqnarray}
Since $\rho$ is an edge state, we have $\varepsilon>0.$ This implies 
that $\tr[W'\rho]<0,$ thus, $\rho$ is detected. 
Also, it is clear that the difficult part of this construction is the 
minimization procedure in Eq.~(\ref{witnessmin}) -- which is just of 
the type of Eq.~(\ref{optigleichung}). 

This method can also be used to obtain a finer witness from a 
given one $\tilde{W}$, i.e., a witness that detects the 
same states as $\tilde{W}$ -- and more. If 
$\tilde{\varepsilon}=\min_{\ket{a,b}} \tr[\ketbra{a,b}\tilde{W}]>0$
then $\tilde{W}-\tilde{\varepsilon}\Eins$ is a finer witness
than $\tilde{W}$. This can also be applied in the scenario where 
only a restricted set of observables is available, since the 
observable $\Eins$ is always accessible. 
In practical situations, given a particular
implementation of a quantum key distribution (QKD)
scheme, it is sufficient to obtain one
relevant entanglement witness as a first step towards the
demonstration of the feasibility of the scheme. Here is where the
method presented in this section can be used. Although this
method requires, as a starting point, to have already a valid
entanglement witness for the given QKD protocol, note that this
operator does not need to be an entanglement witness in the  
strict sense, but can be a positive operator from the restricted
set which is more easy to characterize than an entanglement
witnesses \cite{Full}. Moreover, during several steps of the
method, better entanglement witnesses can be obtained 
from it, belonging to 
the same restricted set.

The construction of the witness above can also be used for
multi-partite PPT witnesses \cite{gbes}. The other 
approaches for 
the construction of entanglement witnesses also need similar 
optimization processes \cite{constructW}. For the sake of 
generality, we formulate 
the optimization strategies directly in the multi-partite setting as in Eq.\ 
(\ref{optigleichung}). 
For a given entanglement witness $W=W^\dagger$, 
the optimization problem looks as follows: 
the aim is to solve the problem
\begin{eqnarray}
	\text{minimize }&& x,\\
	\text{subject to } && x\geq \tr[W P],\nonumber \\
	&&\tr[ \tr_{I\backslash j} [ P]^2]=1,\text{ for all } j\in I, \nonumber\\
	&& \tr[ \tr_{I\backslash j} [ P]^3]=1 ,  
	\text{ for all } j\in I.\nonumber
\end{eqnarray}
Again, for multi-party 
qubit systems this can be written as a polynomially constrained problem
with polynomials of degree two.

\subsection{Estimating the geometric entanglement to 
quantify multi-particle entanglement}

The same tools can be used in order to quantify multi-particle entanglement
for pure quantum states. Needless to say, the question of quantifying 
multi-particle entanglement is much more involved that the analogous
question in the bi-partite setting: in the bi-partite setting, the degree 
of 
entanglement of pure states can be uniquely quantified in terms of
the entropy of entanglement. Any pure state can be asymptotically
reversibly transformed into any other, the achievable rates being 
given by just this measure of entanglement. In this sense, any bi-partite
entanglement of pure states is essentially equivalent to that of the
maximally entangled pair of qubits, which forms the so-called
minimal reversible entanglement-generating set  (MREGS) \cite{MREGS}. 
 The situation is very different in the
multi-partite case, where the MREGS have not even been identified 
for three-qubit systems, let alone for more general settings \cite{MREGS2}.
In the view of this fact, several more pragmatic (and inequivalent)
measures of entanglement
have been proposed, reasonably grasping the degree of multi-particle 
entanglement \cite{Linden,Wei,Multi}. 
To evaluate these quantities typically amounts
to solving a computationally hard problem. 

One of the 
reasonable quantities to quantify multi-particle entanglement is 
the geometric measure of entanglement \cite{Shimony,Linden,Wei}: 
for a given 
state vector $|\psi\rangle\in {\cal H}_1\otimes ...\otimes {\cal H}_N$,
essentially, entanglement is then quantified in terms
of the solution of the maximization problem
\begin{eqnarray}
	\Lambda^2= \text{max}    | \langle \psi |\phi\rangle |^2,
\end{eqnarray}
such that the geometric measure of 
entanglement becomes
\begin{equation}
	E(|\psi\rangle\langle\psi|)= 1-\Lambda^2.
\end{equation}
The maximization is performed
over all state vectors $|\phi\rangle$ which are products with
respect to all subsystems.
Setting $\rho=|\psi\rangle\langle \psi|$ and
\begin{equation}
	P=|\phi\rangle\langle\phi|= |\phi_1\rangle\langle\phi_1|
	\otimes ...\otimes 
	|\phi_N\rangle\langle\phi_N|	,
\end{equation}	
we arrive at
\begin{eqnarray}
	\text{minimize }&& t,\\
	\text{subject to }
	&&\tr [P \rho]+t\geq 1,\nonumber\\	
	&&\tr[ \tr_{I\backslash j} [ P]^2]=1, \text{ for all } j\in I, \nonumber 
\\
	&& \tr[ \tr_{I\backslash j} [ P]^3] =1 ,  
	\text{ for all } j\in I,\nonumber
\end{eqnarray}
which is the same optimization as in the previous subsection, except from 
one line in the
list of constraints.

\subsection{Entanglement 
witnesses based on second moments and entropic criteria}

In this subsection we will consider again entanglement witnesses, but not
in the original sense, which involve only expectation
values of Hermitean operators. It is also possible to introduce
nonlinear functionals with similar properties: these are 
entanglement witnesses based on second moments,
on variances of observables. 
Such entanglement criteria based on 
second moments are very popular in the study of 
infinite-dimensional quantum systems having canonical 
coordinates \cite{Gaussian}.
There, to measure arbitrary observables is often by far unfeasible, 
whereas the estimation of second moments of canonical coordinates
is very accessible. In optical systems, the appropriate 
measurements are available in homodyne detection. 
Similarly,
one may also for finite-dimensional systems look at variances rather
than at first moments themselves \cite{Hof}. 
In Ref.\ \cite{Guehne},
second order witnesses related to variances of operators
were constructed, and the relation to entanglement criteria
for continuous variable systems based on second moments was shown. 
The advantage in a practical context is that one
specifies some observables which are the 
most accessible, and tests whether the
obtained second moments are consistent with a separable state. The 
application of such a
test always requires the solution of an optimization problem as pointed 
out below.

Let us specify a set of observables $M_1,...,M_K$. Then we can define
the real symmetric $K\times K$
covariance matrix $\gamma_\rho$
of a state $\rho$ associated with these observables
as
\begin{eqnarray}
	(\gamma_\rho)_{k,l} &=&
	(\tr[ M_k M_l \rho ] + \tr[ M_l M_k \rho ])/2\nonumber\\
	& -&  \tr[M_k \rho ]\, \tr [M_l  \rho],
\end{eqnarray}
with $k,l=1,...,K$. 
This is completely analogous to the familiar covariance 
matrix
of systems with canonical coordinates. 
Then, it turns out that -- in the previous notation --
any fully separable state $\rho$
has the property that there exist states 
\begin{equation}
\rho^{(i)}_1,...,\rho^{(i)}_N, \,\,i=1,...,n, 
\end{equation}
$n=\prod_{j=1}^N d_j^2$,
and probability distributions
$\{p_i\}_i$ such that 
\begin{equation}\label{Second}
	\gamma_\rho \geq \sum_i p_i \gamma_{\rho^{(i)}_1\otimes ...
	\otimes \rho^{(i)}_N}.
\end{equation}
So one would fix those 
observables that are
the most accessible, and estimate the appropriate second moments.
This would yield an estimate of the elements of the covariance matrix
to some accuracy. Then, the question that arises is: do states
and probability distributions exist that satisfy Ineq.\ (\ref{Second})?
If not, we can conclude that the state must have been entangled. It
is important to note that this judgement is not based on the knowledge
of the entire state, but only on the knowledge 
of the covariance matrix with respect to a previously selected set of 
observables.
This is  a problem that can be cast into 
a feasibility problem, again in the form that we envision. 
As it 
is a feasibility problem, the objective function can be set to zero.
This can be written as follows,
\begin{eqnarray}
\text{minimize}&& 0\\
\text{subject to } && Q- \sum_{i=1}^n p_i Q^{(i)}=0,\nonumber\\ 
&& Q^{(i)}_{k,l}= 
	(\tr[ M_k M_l  P^{(i)}] + \tr[ M_l M_k P^{(i)} ])/2\nonumber\\
	 && \hspace{.7cm} -  \tr[M_k P^{(i)} ]\, \tr [M_l  P^{(i)}],\nonumber\\
	&&\tr[ \tr_{I\backslash j} [P^{(i)}]^2 ] = (\tr [P^{(i)} ])^2 
\,\nonumber\\
	&&\hspace{1cm}\text{ for all } i=1,...,n,\, j\in I,\nonumber\\
		&&\tr[ \tr_{I\backslash j} [P^{(i)}]^3 ] = (\tr [P^{(i)} ])^3 
\,\nonumber\\
		&&\hspace{1cm}\text{ for all } i=1,...,n,\, j\in I,\nonumber\\
	&& \sum_{i=1}^n \tr[P^{(i)}]=1,\nonumber\\
	&&\gamma_\rho - Q\geq 0.\nonumber
\end{eqnarray}
In each step of the  hierarchies of relaxations, 
we can assess whether there is a feasible solution or not. 
If there is no feasible solution in some step, we can conclude that 
the state is entangled, and multi-particle entanglement is hence
detected.
This is now a problem which is still a combination of a polynomially
constrained problem, together with a semi-definite constraint.
Here, two strategies may be applied: 

On the one hand,
one can keep the semi-definite constraint, and can proceed as pointed out in Subsection A. 
This means that one can in each step assert that the state was entangled, or one has
to go one step further. This is computationally cheaper, but comes at the price of 
losing asymptotic completeness. For practical purposes, however, this 
method is expected to be the method of choice; in particular in the light
of the fact that for a given set of observables, typically
not every entangled state is anyway detected by
the entanglement witness based on second 
moments. 


The other strategy, on the other hand, 
is to formulate $\gamma_\rho - Q\geq 0$ as a set of 
polynomial constraints. A Hermitean matrix is positive if and only if 
the determinants of all its submatrices are positive
(see also Ref.\ \cite{Full}). 
This gives rise to a set of polynomial constraints, for which the 
relaxations can be applied, leading to an asymptotically complete hierarchy
of tests. This comes at the price of being computationally more expensive.

In view of Refs.\ \cite{Hof,Guehne}, it is useful to employ arguments
along the following line: if we can show that no fully 
separable state can have the image that we estimate in an experiment, we 
can assert that the state must have been entangled. This observation can, 
while being fairly 
obvious, still be practically very relevant. For example, we may for any
observable $M=M^\dagger$ look at the minimum of 
second moments that are consistent with a separable state, i.e., 
the solution of
\begin{eqnarray}
	\text{minimize}&& \tr[M^2 P]- \tr[M P]^2,\\
	\text{subject to } && P \text{ is fully separable},
\end{eqnarray}
and use this as a criterion for detecting entangled states. 
This gives rise to the optimization problem, 
\begin{eqnarray}
	\text{minimize}&& x,\\
	\text{subject to} && 
	x \geq \tr[M^2 P]- \tr[M P]^2  ,\nonumber\\
	&&P-\sum_{i=1}^n P^{(i)} =0,\nonumber\\
	&&\tr[ \tr_{I\backslash j} [P^{(i)}]^2 ] =  (\tr [P^{(i)} ])^2,  
\,\nonumber\\
		&&\hspace*{1cm}	\text{ for all } i=1,...,n,\, j\in I . \nonumber\\
		&&  \tr[ \tr_{I\backslash j} [P^{(i)}]^3 ] =  (\tr 
[P^{(i)} ])^3,  \,\nonumber\\
		&&\hspace*{1cm}
		\text{ for all } i=1,...,n,\, j\in I . \nonumber\\
	&& \sum_{i=1}^n \tr[P^{(i)}]=1.  \nonumber
\end{eqnarray}

It should be clear at this point 
that the same method can be
used for entanglement criteria based on linear entropies, that is,
$p$-norms for $p=2$ (see, in the rich literature on the subject,
e.g., 
Refs.\ \cite{Linear}). 
 For any expression that is linear in
the linear 
entropies of the whole state $\rho$, 
\begin{equation}
	\|\rho\|_2= 
\tr[\rho^2]
\end{equation}
of a multi-partite
system and  in the linear 
entropies of the
reductions 
\begin{equation}
\|  \tr_{j\backslash 
I}[\rho] \|_2 
= \tr[(
\tr_{j\backslash I}[\rho])^2], \,\, j=1,...,n,
\end{equation}
one can in the same manner
find the largest value 
consistent with a separable state.
Any state that delivers a larger value 
is then clearly entangled.
In practical considerations, these linear 
entropies can be estimated in a fairly feasible manner
\cite{Linear}, for example when assessing entanglement in 
Bose-Hubbard-type models. 

\subsection{Maximal output purities of quantum channels}

Similar arguments, it shall finally be briefly 
discussed, can immediately 
be applied to assess minimal  
output purities of channels 
\begin{equation}
	\rho\longmapsto {\cal E}(\rho)= \sum_{i=1}^k R_i \rho R_i^\dagger
\end{equation}
where $\sum_{i=1}^k R_i^\dagger R_i=\id$,
with respect to $p$-norms for $p=2$ (and other integer $p$), see, e.g.,
Refs.\ \cite{Channel}. One may then investigate the
maximal output purity
\begin{equation}
	\nu_2({\cal E})=	\max_{\rho} || {\cal E}(\rho)\|_2 ,
\end{equation}
where it does not constitute a restriction of generality to
maximize not over all states $\rho$, but merely 
over all pure states (compare also Ref.\ \cite{Global}). The central question here
is to see whether this quantity is multiplicative in general. That is
whether generally
\begin{equation}
	\nu_2({\cal E}_1\otimes {\cal E}_2) = \nu_2({\cal E}_1)
	 \nu_2({\cal E}_2)
\end{equation}
holds, which means that it is never an advantage to allow
for entangled inputs when maximizing the output purity.
In the previously used language, this 
optimization problem can be written as follows,
\begin{eqnarray}
	\text{maximize}&& x,\\
	\text{subject to} && 
	x\leq \tr[ (\sum_{i=1}^k R_i P 
	R_i^\dagger )(\sum_{j=1}^k R_j P R_j^\dagger )]  ,\nonumber\\
	&& \tr[P^2]=1,\nonumber\\
	&& \tr[P^3]=1,\nonumber
\end{eqnarray}
as a polynomially constrained problem with polynomials of degree three.

\section{Complete 
hierarchies of relaxations to approximate the solutions}
\label{boyrelax}

We will now state how the theory of relaxations can be applied
to the described problems in entanglement theory.  
In all of the above cases (except in Subsection D), 
we obtained an optimization problem 
of the following structure: for
$x\in \rr^t$,
\begin{eqnarray}
	\text{minimize} && c^T x\\
	\text{subject to} 
	&& g_l(x) \geq 0, \nonumber
\end{eqnarray}
for $l=1,...,L$, the global optimum value being denoted as $p^\ast$, where
$g_l$ is a polynomial of at most degree three.
The constraint set given by
\begin{equation}
	{\cal M}=\{ x\in \rr^t:  g_l(x)  \geq 0,\, l=1,...,L\}
\end{equation}
is not a convex set. We may however apply Lasserre's method of 
semi-definite relaxations to treat this part, see Appendix A. 
This will yield a sequence of
semi-definite programs, labeled with an index 
$h=h_{\rm min},h_{\rm min}+1,...$, such that each
of the efficiently solvable steps yields an approximation of the original
problem. The minimal step $h_{\min}$ is 1 if the highest
degree of the constraint polynomials is $2$ and $h_{\rm min}=2$
if constraints of degree $3$ are required.
The case $h=h_{\rm min}$ is the first semi-definite relaxation 
leading to the first approximation, $h=h_{\rm min}+1$ is the 
second, and so on. Often, in practice the global optimum is 
already achieved after a small number of steps in the hierarchy.

\subsection{Semi-definite relaxations}

Instead of considering an optimization problem in $x\in \rr^t$, this
is turned into an optimization problem in a larger real vector $y\in 
\rr^{D_{2h}}$, 
$D_{2h}$ being a natural number defined in 
Appendix A. This larger dimension is due to the uplifting procedure used 
in 
Lasserre's 
method \cite{Lasserre,Lasserre2} 
for approximating the
quadratic part of our problems, and goes back to work in Ref.\ \cite{SA90}.
For each instance of the hierarchy of semi-definite programs, the
objective function will be the same, but uplifted, namely
\begin{equation}
	y\longmapsto d^T y,
\end{equation}
where
\begin{equation}
	d^T=(0,c_1,...,c_t,0,...,0),
\end{equation}
with $c\in \rr^t$ being defined as above. 
Lasserre's method now gives rise to a
sequence of semi-definite programs approximating the solutions of
\begin{eqnarray}\label{PolProb}
	\text{minimize} && c^T x\\
	\text{subject to} && g_l(x) \geq 0, \nonumber
\end{eqnarray}
for $l=1,...,L$ in the following form:
For $h=h_{\rm min},h_{\rm min}+1, ...$, each instance is of the form
\begin{eqnarray}\label{Originalproblem}
	\text{minimize}&& d^T y,\\
	\text{subject to} && F^{[h]}(y) \geq 0,\nonumber\\
	&& G^{[h]}_l(y)\geq 0,\,l=1,...,L \nonumber
\end{eqnarray}
with matrices $F^{[h]}(y)$ and $G^{[h]}_l(y)$ that are linear in the 
elements of $y$ that increase in 
dimension 
with increasing $h$ (see Appendix A). This method is based on recent 
results in real algebraic geometry, see also Ref.\ \cite{Par}.

In Ref.\ \cite{Lasserre} convergence to the solution of 
(\ref{PolProb}) is guaranteed
if certain conditions are satisfied:  Convergence in the limit 
$h\rightarrow \infty$ is
guaranteed if
there exist polynomials, $u_0,u_1,...,u_L$, all sums of squares, 
such that the set
\begin{equation}\label{cond}
	\{ x\in \rr^t : u_0(x) + \sum_{l=1}^L 
	u_l(x) g_l(x)\geq 0\}
\end{equation}
is compact. 
This is, however, 
the case in all of the specific situations from 
entanglement theory considered above. 
The set in Eq.\ (\ref{cond}) 
is compact if there exists an $l\in\{1,...,L\}$  such that the set
\begin{equation}\label{sc}
	\{ x\in \rr^t :   g_l(x)\geq 0\}
\end{equation}
is compact. In each of the discussed cases, we find that due
to the linear constraints 
incorporating the trace requirement 
and the 
quadratic constraints coming from the purity of the 
reduced states, there exists an $a>0$ such that 
$a^2-\| x\|^2\geq 0$ for all $x\in {\cal M}$. This follows from the fact that 
for each of the involved matrices, the trace is bounded from above, and 
positivity of the matrices enforces boundedness of all elements.
Hence, to ensure
asymptotic completeness, we may add the constraint 
$g_{L+1}(x)= a^2-\| x\|^2\geq 0$ to the list of quadratic 
constraints, such 
that the condition in Eq.\ (\ref{sc}) is certainly satisfied.
Hence, one can conclude that 
\begin{eqnarray}
	\min_{y\in {\cal M}^{[h]}} d^T y\rightarrow \min_{x\in {\cal M}}  c^T x
\end{eqnarray}
for $h\rightarrow\infty$, and for 
\begin{eqnarray}
	{\cal M}^{[h]}&=& \{y\in\rr^{D_{2h}}:  F^{[h]}(y) \geq 0,\,\nonumber\\
	&&\hspace*{1cm}  G^{[h]}_l(y)\geq 0, \,l=1,...,L+1 \}.
\end{eqnarray}
This is not only meant as a numerical procedure: instead, as each step is 
an
analytically accessible semi-definite program, in each step one may
assess the approximations with analytical means. Moreover, symmetries
of the involved states under certain groups can be carried over to 
symmetries
in the Hermitean matrices in the semi-definite programs, similarly to the 
strategy
employed in Ref.\ \cite{Rains} for semi-definite programs, and in 
Ref.\ \cite{ConvexButNotSemidef}
for convex but not semi-definite programs.

In Subsection D of Section \ref{sec:2}, we encountered an additional 
semi-definite
constraint. Then, Lasserre's method may be applied using polynomials of 
higher order, as described above. Or, one may combine the semi-definite 
relaxations with the semi-definite constraint itself. This gives rise to a
hierarchy of sufficient tests, without the property of asymptotic 
completeness. To see how they can be combined, let us consider an 
additional semi-definite constraint such as 
$\gamma_\rho - Q\geq 0$. 
In terms of the $y\in \rr^{D_{2h}}$, we 
have the feasible set of the additional
semi-definite constraint
\begin{equation}
	{\cal F} =\{ y\in\rr^{D_{2h}}: F_0 + \sum_{s=1}^t y_{s+1} F_s\geq 0\},
\end{equation}
with appropriate matrices $F_0,...,F_t$.
Therefore, we can write the full hierarchy of semi-definite programs as
\begin{eqnarray}
	\text{minimize}&& d^T y,\\
	\text{subject to} && F^{[h]}(y) \geq 0,\nonumber\\
	&& G^{[h]}_l(y) \geq 0,\,l=1,...,L \nonumber\\
	&& F_0 + \sum_{s=1}^t y_{s+1} F_s\geq 0\nonumber,
\end{eqnarray}
$h=h_{\text{min}},h_{\text{min}+1},
 ... $ being the label of the element of the hierarchy. The 
projection of the feasible sets ${\cal M}^{[h]}$ onto the plane of
first order moments, i.e., onto the 
plane 
\begin{equation}
	\{
	y\in \rr^{D_{2h}}: y=(0,y_2,...,y_{t+1},0,...,0)\},
\end{equation}
conceived as a subset of $\rr^t$, converges (pointwise) 
to the convex hull of ${\cal M}$ \cite{Lasserre,Lasserre2,Kojima}. 
Therefore, we have that
\begin{eqnarray}
	\min_{y\in {\cal M}^{[h]}\cap {\cal F}} d^T y \leq 
 	p^\ast
\end{eqnarray}
for all $h\rightarrow\infty$. Moreover, $\min_{y\in {\cal M}^{[h]}\cap {\cal 
F}} d^T y$
is a monotone increasing sequence in $h$, 
such that the sufficient criteria become 
more powerful with an increasing order of the hierarchy.

\subsection{Size of the relaxations}

A relevant issue is  how large the semi-definite
relaxations are in each step of the hierarchy. In the worst-case
scenario, where the polynomial constraints is a polynomial involving all
basis elements of the basis of polynomials of the respective degree, 
one obtains the subsequent sizes of the relaxation matrices.
The matrix $F^{[h]}$ is of 
dimension 
$D_h\times D_h$ (for a definition of $D_h$, see Appendix A).  
As a formula for $D_h$ we arrive at 
\begin{equation}
	D_{h}=\sum_{k=0}^{h}{t+k-1 \choose k}.
        \label{size}
\end{equation}
For example, 
\begin{equation}
	D_2= 1+t+ \frac{t(t+1)}{2}.
\end{equation}
In the number of variables $t$, this is a manifestly 
polynomial expression. 
In step $h$ the vector $y$ is of the length $D_{2h}$.
Notably, in each of 
the steps, the effort
of a numerical solution of the associated semi-definite program
is polynomial in the dimension of the matrices 
\cite{Semi}. Hence,
each problem can be solved in an efficient manner. 

In terms of the step $h$ in the hierarchy, it turns out that the scaling is
also polynomial. Approximating the above sum by an integral expression, we arrive at
\begin{equation}
	D_h=O(h^t).
\end{equation}
That is, for a fixed number of variables (which is the setting considered 
here), the size of the vector of the objective variables increases also 
only polynomially in the step $h$ in the hierarchy. 
Moreover, in many 
small and medium size problems, 
the programm detects optimal solutions in the 
first iteration steps at relatively low
computational cost \cite{gpmanual}. Also, the sparsity of 
the moment matrices may be exploited. The issue of computational effort 
will be discussed in more detail elsewhere.
Another point of interest is that it is 
possible in some cases to trade in a lower number
of variables $t$ for a higher lowest relaxation
step $h_{\rm min}$, as in the examples in the
subsequent section. In some cases, this might simplify
 the problem, as in our example in the next section.	

\section{Numerical Examples}

In this section  we  present some numerical examples, in order to show
that the approach is also feasible in practice. We will provide three 
examples, 
two for the geometric measure of entanglement and one for the construction
of entanglement witnesses for bound entangled three-qubit states. 

\subsection{Geometric measure for three-qubit states}

Let us start with the calculation of the geometric measure of entanglement
for three-qubit states. As we have shown in Section \ref{sec:2} 
the computation of the geometric measure of entanglement for a 
given pure three-qubit state vector
$\ket{\psi}$ requires essentially the calculation of 
\be
\Lambda^2=\max_{\ket{a,b,c}}|\braket{a,b,c}{\psi}|^2
\ee
As already mentioned above, we use here a different parametrization 
than the general one described in Section II.
In terms of the Pauli matrices forming a basis of Hermitean 
matrices, we can write
\bea
\ketbra{\psi}&=&\frac{1}{8}\sum_{i,j,k=0}^3 \lambda_{ijk} (\sigma_i 
	\otimes \sigma_j \otimes \sigma_k),
\\
\ketbra{a,b,c}&=&\frac{1}{8}\sum_{i,j,k=0}^3  a_i b_j c_k (\sigma_i 
	\otimes \sigma_j \otimes \sigma_k),
\eea
where $\lambda_{000}=a_0=b_0=c_0=1.$ The coefficients $\lambda_{ijk}$, 
$i,j,k=0,...,3$, 
are determined from the known state vector $\ket{\psi}$. We have to impose 
constraints
that guarantee that $\rho_A$ is a pure state on the 
coefficients $(a_1,a_2,a_3)$ describing the state 
$\rho_A=  \sum_{i=0}^3 a_i \sigma_i/2 $ (and  similarly 
$(b_1,b_2,b_3)$ and $(c_1,c_2,c_3)$). 
We have seen before that 
for qubit systems, instead of requiring $\tr[\rho_A^2]=1$ and 
$\tr[\rho_A^3]=1$, we 
may alternatively merely require that $\tr[\rho_A]=1$ and $\tr[\rho_A^2]=1$
(where $\tr[\rho_A]=1$ is already a consequence of the parametrization). 
So we arrive at the optimization problem
\begin{eqnarray}\label{newlab}
\text{minimize}_{a_i,b_j,c_k} && \frac{1}{8}  \sum_{i,j,k=0}^3 
\lambda_{ijk} a_i  b_j c_k, \\
\text{subject to} && 
a_1^2+a_2^2+a_3^2=1,
\nonumber \\
&& b_1^2+b_2^2+b_3^2=1,
\nonumber \\
&& c_1^2+c_2^2+c_3^2=1.\nonumber
\eea
This polynomial optimization problem can be solved
with the help of  
Lasserre's method, see Appendix A. 
For these calculations the package GloptiPoly 
\cite{gpmanual} based on SeDuMi \cite{SeDuMi}
is freely available, and we have used it 
for our calculations. The package GloptiPoly has a number of
desirable features, in particular, it provides a 
certificate for global optimality. 

Note that with this parametrization, the number
of objective 
variables is $4N$,  $N$ being the number
of qubits, in contrast to $4^{N}$ parameters
which are necessary to parametrize a general
$N$ qubit state as decribed in Section II.A. 
From Eq.\ (\ref{newlab}) 
it is clear that 
the objective function will be a polynomial of 
degree $N$ which increases $h_{\rm min}$,
see Appendix A.

First, we present a nontrivial example for the calculation of
the geometric measure of entanglement, in a case 
where its value is already known. In this way we can 
test our methods. We aim at computing the geometric 
measure of entanglement for state vectors of the form
\be
\ket{\psi(s)}= \sqrt{s}\ket{W}+ \sqrt{1-s}\ket{\tilde{W}},
\label{weistate}
\ee
 $s\in [0,1]$,
where $\ket{W}$ and  $\ket{\tilde{W}}$ are state vectors of
three-qubit W states 
\cite{duer} in different bases,
\bea
\ket{W}&=& (\ket{001}+\ket{010}+\ket{100})/\sqrt{3},
\\
\ket{\tilde W}&=& (\ket{011}+\ket{101}+\ket{110})/\sqrt{3}.
\eea
For the geometric measure of entanglement of $\ket{\psi(s)}$ a formula 
has been developed in Ref.\ 
\cite{Wei}, exploiting the permutation symmetry of the
states. 
 The comparison between the theoretical 
value and the numerical calculation using Lasserre's method for $h=2$
is shown in Fig.~\ref{fig:wei}. Details of the performance 
are summarized in Table \ref{tabl}.
The results indicate clearly the usefulness of 
the presented approach. As a matter of fact, this is a case where already
a very small number of steps in the hierarchy detects the global optimum, 
as is 
typical for this relaxation, as has been pointed out in Ref.\ 
\cite{gpmanual}, based on
numerical experiments.
\begin{figure}[h]
   \includegraphics[width=\columnwidth]{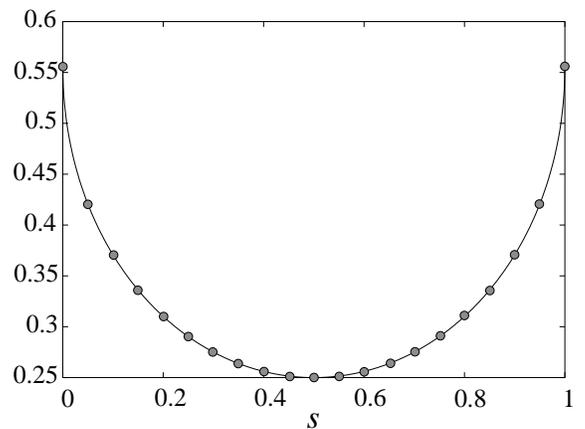}
    \caption{The numerical values of the geometric measure of entanglement $E$
	of the family of states of Eq.\ (\ref{weistate}), plotted
	on top of the analytical values of Ref.\ \cite{Wei}.}
 \label{fig:wei}
\end{figure}

\begin{table}
\begin{tabular}{|c|c|c|c|c|}
\hline\hline
    Subsection &Relaxation $h$ & $\#$ variables & $\dim(y)$ & CPU time\\
\hline
A& 2 & 9 & 714 & 10.92 $s$\\
B&  2 & 12 & 1819 & 103.97 $s$\\
C&  2 & 9 & 714 & 6.14 $s$\\
\hline\hline	
\end{tabular}
\caption{Details of the relaxations in the three
numerical examples discussed above for one point
of each example. The provided CPU time refers 
to a machine with a Intel Xeon Processor, 2.2 GHz,  1GB Ram, using
GloptiPoly 2.2e \cite{gpmanual}, SeDuMi 1.05 \cite{SeDuMi}, and MatLab
6.5.1.199709 (release 13).
In all cases $h=h_{\rm min}=2$, so that the
result was obtained after the first relaxation step. }\label{tabl}
\end{table}

\subsection{Geometric measure for 4-qubit states}
We calculate the geometric measure of entanglement
also for the following one parameter family of state 
vectors
\be
	\ket{\psi_{4}(p)}=\sqrt{p}\ket{\text{GHZ}'}
	-\sqrt{1-p}\, 
	\ket{\psi^{+}}\otimes\ket{\psi^{+}},
	\label{psi4state}
\ee
where 
\begin{equation}
\ket{\text{GHZ}'}=(\ket{0011}+\ket{1100})/ \sqrt{2},
\end{equation}
$\ket{\psi^{+}}=(\ket{01}+\ket{10})/\sqrt{2}$, and 
$p\in[0,1]$.
The state vector $\ket{\psi_{4}(2/3)}$ corresponds to 
the 4-qubit singlet state, i.e., the state vector satisfying 
\begin{equation}
U^{\otimes 4}\ket{\psi}=\ket{\psi}
\end{equation}
for all unitary $U$  \cite{kempe}. For the two individual states
in the above 
superpositions in Eq.\ (\ref{psi4state}), 
the geometric measure can be directly evaluated
\cite{Wei}: For $p=1$ we find $\Lambda^2=1/2$, and for 
$p=0$ we obtain 
$\Lambda^2=1/4$ from $\Lambda^2_{\psi^{+}}=1/2$.
The numerical results for the geometric measure of entanglement
for other values of $p$ are plotted in 
Fig.~\ref{fig:psi4}.
\begin{figure}[h]
   \includegraphics[width=\columnwidth]{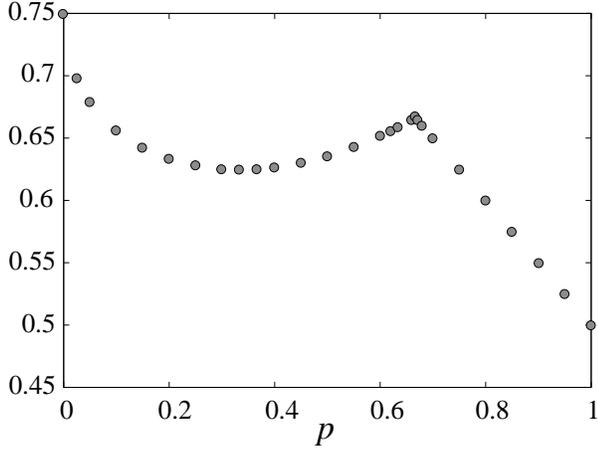}
    \caption{The numerical values of the geometric measure of entanglement
	$E$
    of the family of states of Eq.\ (\ref{psi4state}).}
 \label{fig:psi4} 
\end{figure}
It is interesting to note that at the singlet value $p=2/3$, the behavior 
of
the geometric measure changes. From there 
up to $p=1$ the optimum is attained for 
the choices $\ket{0011}$ or $\ket{1100}$ of the product state which 
gives rise to the linear behavior.

The family of states specified in Eq.\ 
(\ref{psi4state}) is invariant under the 
exchange $(AB)\leftrightarrow(CD)$.
Because of this symmetry, one may without loss of generality
assume that the product state vector 
leading to the maximal value of $\Lambda^2$ is given by
$\ket{\phi_{1},\phi_{2},\phi_{1},\phi_{2}}$, where
$\ket{\phi_{1,2}}=e^{i \chi_{1,2}  }\cos{\theta_{1,2}}\ket{0}
+e^{i \eta_{1,2}  } \sin{\theta_{1,2}}\ket{1}$, where
the optimal phases can be shown to be
$\chi_1=\chi_2=\eta_1=\eta_2=0$. This gives rises to an
optimization problem with polynomial constraints
with only 4 variables which can be solved exactly by GloptiPoly. 
The results coincide with the results above.

\subsection{Witness for 3-qubit PPT entangled states}

Employing the same strategy, we would like to calculate 
the value of $\varepsilon$ as defined in Section \ref{sec:2}
for the family of witnesses constructed for the 
PPT (bound) entangled states
\bea
\rho &=&  \label{BES} 
\Big(a\KetBra{001}+b\KetBra{010}+c\KetBra{011}\\
&&	+\frac{1}{c}\KetBra{100}+\frac{1}{b}\KetBra{101}
		+\frac{1}{a}\KetBra{110}\nonumber\\& + &2\KetBra{\text{GHZ}}\Big)
		/(2+a+b+c+1/a+1/b+1/c)\nonumber 
\nonumber	
\eea
where $a,b,c>0$,  $ab\neq c$,
and 
\begin{equation}
|\text{GHZ}\rangle=(|000\rangle + |111\rangle)/\sqrt{2}.
\end{equation}
In Ref.\ \cite{gbes}, upper bounds for the values of $\varepsilon$ 
were obtained by using a multi-variable minimization routine for 
the parameter range $a=b=1/c\in(0,1)$. The minimization
over the product states has to be performed 
with respect to \cite{gbes}
\begin{eqnarray}
\bar{W}&=&\frac{1}{2}\Big(\KetBra{000}+\KetBra{111}\Big)  
\label{prewitness} \\ 
&+&\frac{1}{1+c^{2}}\Big(c^2\KetBra{100}+\KetBra{011}\Big) \nonumber\\
&   + &\frac{1}{1+b^{2}}\Big(\KetBra{010}+b^{2}\KetBra{101}\Big)
    \nonumber\\
    & +& \frac{1}{1+a^{2}}\Big(\KetBra{001}+a^{2}\KetBra{110}\Big) 
    \nonumber\\
    & -&\Big(\frac{1}{2}+\frac{c}{1+c^{2}}+\frac{b}{1+b^{2}}
    +\frac{a}{1+a^{2}}\Big) \nonumber\\
    &\times &\Big(\ket{000}\bra{111}+\ket{111}\bra{000}\Big).\nonumber
\end{eqnarray}
The numerical results are plotted in Fig.~(\ref{fig:bound}). Again, the 
global optimum is achieved, and the found values agree with
the values found in Ref.\ \cite{gbes}. For details concerning the 
relaxations, see Table \ref{tabl}.
\begin{figure}[h]
   \includegraphics[width=\columnwidth]{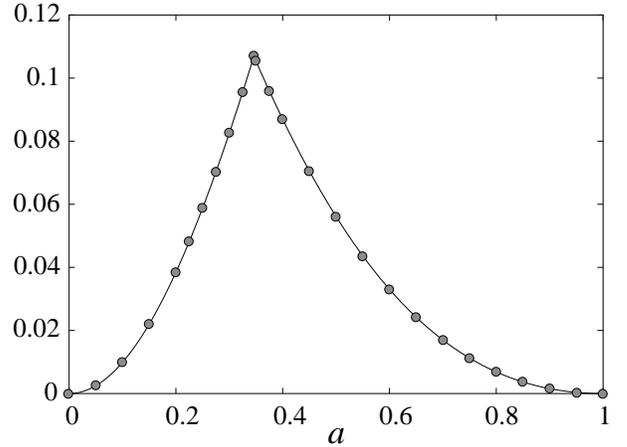}
    \caption{The numerical values of $\varepsilon$ for 
	$\bar{W}$ of Eq.~(\ref{prewitness}) plotted on top of
	the results of Ref.\ \cite{gbes}.}
 \label{fig:bound}
\end{figure}

\section{Summary and outlook}

In this paper, we have considered 
several problems in entanglement theory with
the tools and language of convex optimization. The central point of the 
paper was that many problems, where a minimization over pure 
product vectors is required, can 
be written as
instances of certain optimization problems involving polynomial 
constraints of degree two or three, 
or with additional 
semi-definite constraints. For such polynomially constrained
problems, which are generally 
instances of non-convex optimization problems, 
hierarchies of semi-definite relaxations can be found. 
In this sense, one additional intention of this paper
is to communicate these recently achieved results 
in the theory of relaxations 
and to show that they can be fruitfully applied in the 
quantum information context.
One arrives at hierarchies
of more and more refined tests detecting entangled or separable states, or 
better and better
lower bounds to optimization problems. In all instances, recently achieved 
known results from semi-algebraic geometry 
guarantee that asymptotically, the achieved minimum is indeed
approaching the globally optimal one. In this sense, the statements are 
similar in spirit with 
yet more versatile than 
the ones presented in Refs.\ \cite{Doherty}. Moreover, we have seen that the
size of the optimization problems to be solved in each test grows
polynomially with the steps in the hierarchy, and that for small problems, 
often already a small number of steps is required to  find the exact
solution. 

The presented method is on the one hand meant as a numerical method to 
achieve good bounds to problems that are of relevance in the study of 
multi-particle entanglement,
in the construction of entanglement witnesses in the bi-partite and 
multi-partite case, 
in the context of quantum key distribution, and to assess maximal 
output purities. 
On the other hand, 
each instance of the hierarchy delivers
a semi-definite program which is readily accessible with analytical 
methods, and where properties of the
Lagrange-dual can be exploited. It is hoped 
that these techniques shed new light on the  structure of 
optimization problems underlying the questions of entanglement and 
separability of 
several constituents.

Finally -- and shifting perspective to some extent --
it seems worth noticing that very similar techniques
may be expected to 
be useful tools to assess ground state properties of
many-body Hamiltonians. Often variational
approaches deliver already a good approximation to properties
of the true ground state. For example, in the Gutzwiller ansatz for the 
ground state of the Bose-Hubbard model in a lattice one optimizes the 
energy functional over product states with respect to the sites. 
Similar techniques can be used for spin systems and 
expressions for ground states in terms of matrix product states. 
Then, relaxations in the
way discussed above could potentially be applied  for a reasonable size of 
the system. Such studies could complement numerical 
techniques yielding upper bounds, such 
as simulated annealing techniques, 
delivering provable lower bounds for the ground state energy.

\section{Acknowledgments}

We would like to  thank Masakazu Kojima and Christoph Helmberg
for very thoughtful and detailed communication on the
subject of successive relaxation methods, Andrew C.\ Doherty,
Nick Jones, and Arnold Neumaier
for helpful discussions, and Norbert L{\"u}tkenhaus
for very useful comments on the manuscript. 
This work has been supported by 
 the DFG (Schwerpunktprogramm QIV, SPP
1078, Graduiertenkolleg 282, 
and the Emmy Noether Programme) and the 
European Commission 
(QUPRODIS IST-2001-38877 and
Integrated Project  SECOQC). 
This work has benefited from
discussions during one of the A2 consortial meetings, 
funded by the DFG (Schwerpunktprogramm QIV, SPP
1078). We are grateful to 
Andrew C.\ Doherty for letting us know about
their very recent work \cite{NewAndrew}
and for agreeing to simultaneous
posting. 

\section{Appendix A: Lasserre's method}

For completeness,
in this appendix  we briefly sketch the method to construct
sequences of semi-definite relaxations of global optimization 
problems with multi-variate real-valued polynomial objective function
and constraints due to Lasserre \cite{Lasserre}. 
The class of problems is 
of the following form:
\begin{eqnarray}
	\text{minimize} && c^T x,\,\, x\in \rr^t,\\
	\text{subject to} && g_l(x)\geq 0,\,\, l=1,...,L,
\end{eqnarray}	
where $g_1,...,g_L:\rr^t\rightarrow \rr$ are 
real-valued polynomials of degree two or three.
Although we consider only polynomials of degree of at most three, 
it will be convenient to formulate the subsequent sequence of 
semi-definite programs in terms that formally involve higher-order
polynomials. 
For any $r\in \nn$, we consider the basis of polynomials of degree $r$
in the variables $x_1,...,x_t$ as
\begin{equation}\label{thebasis}
(1;x_1,...,x_t; x_1^2, x_1 x_2,...,x_1 x_t;
x_2^2,x_2 x_3,..., x_t^r),
\end{equation}
in this ordering.  The dimension of this basis will be denoted as
$D_r $. For clarity of notation, we will not specify $t$ as an index,
as this will stay the same throughout the procedure.
Any  polynomial of degree of at most $r$
can then be identified with a vector $p\in \rr^{D_r}$.
It is convenient to introduce two labelings, 
connected with each other by a function 
\begin{equation}
f_r:\{1,...,D_r\} \rightarrow
\biggl\{ \alpha=(\alpha_1,...,\alpha_t) : \sum_{s=1}^t \alpha_s \leq r
\biggr
\},
\end{equation}
such that the $i$-th element $z$, $i=1,...,D_r$, of the basis 
given by Eq.\ 
(\ref{thebasis})
is written as 
\begin{equation}
	z=\prod_{i=1}^t x_i^{\alpha_i},
\end{equation}
characterized by $\alpha=(\alpha_1,...,\alpha_t)\in \nn_0^t$. 
Note that for a given $ k\in \nn $ 
there are ${t+k-1 \choose k}$ possible vectors $\alpha$ 
such that 
$\sum_{s=1}^{t}\alpha_{s}=k$. It follows that 
the dimensions $D_{h}$ are given by 
Eq.\ (\ref{size}).

In the following we give the required matrices from Lasserre's
method for general polynomials \cite{Lasserre} and discuss the
cases occuring in the paper explicitly afterwards.
Let $\delta_{l}$ be the degree of the polynomial constraint
$l\in \{1,...,L\}$ 
and $ \lceil \delta_{l}/2 \rceil$ be the smallest integer greater than
or equal to 
$\delta_{l}/2$. We assume that the objective function is linear,
which is no restriction of generality, as other polynomials can always
be incorporated in the constraints as in Section II. 
Then the first possible relaxation step of Lasserre's method
is $h_{\rm min}=\max_{l  } \lceil \delta_{l}/2 \rceil$.
For $h\ge h_{\rm min}$ the matrix $F^{[h]}(y)$ is of 
dimension $D_{h}\times D_{h}$ and linear in a vector
$y\in \rr^{D_{2h}}$,
\begin{equation}\label{ff}
	\bigl[F^{[h]}(y)\bigr]_{i,j}= y_{f_{2h}^{-1}
	( f_{h}(i)+ f_{h}(j)  )}.
\end{equation}
In turn, the matrices
 $G^{[h]}_l(y)$, one for each of
the constraint polynomials, $l=1,...,L$, are of dimension
$D_{\tilde{h}_{l}}\times D_{\tilde{h}_{l}}$,
where $\tilde{h}_{l}= h- \lceil \delta_{l}/2 \rceil$. 
Each polynomial $g_l$ is characterized according to the
above procedure by a vector $v_l$. 
The matrices $G^{[h]}_l(y)$ 
are then defined as
 \begin{equation}\label{g}
	\bigl[
	G^{[h]}_l(y) \bigr]_{i,j}= \sum_\alpha v_{f_{ \delta_{l}   }^{-1}(\alpha)} 
	y_{(f_{   \delta_{l} 
   +2\tilde{h}_{l}}^{-1}( f_{\tilde{h}_{l}}(i) 
	+ f_{\tilde{h}_{l}}(j)) + \alpha)}.
\end{equation}
For qubits, $h_{\rm min}=1$, because the maximal
degree of the constraint polynomials is $2$.
For higher dimensional systems, the highest occuring 
order is $3$ due to the positivity constraints.
In this case, $h_{\rm min}=2$.


\begin{thebibliography}{99}
%
\bibitem{Gurvits}
L.\ Gurvits, Annual ACM Symposium on Theory of 
Computing,                                                                                                                                                                                                                    
Proceedings of the thirty-fifth ACM symposium on theory of computing,
San Diego, CA, USA, ACM Press, June 9-11   (2003).
%
\bibitem{Wer89}
R.F.\ Werner, Phys.\ Rev.\ A {\bf 40}, 4277 (1989).
%
\bibitem{Separability}
S.L.\ Woronowicz, Rep.\ Math.\ Phys.\ {\bf 10}, 165 (1976);
A.\ Peres, Phys.\ Rev.\ Lett.\ {\bf 77}, 1413 (1996);
M.\ Horodecki, P.\ Horodecki, and R.\ Horodecki, Phys.\ Lett.\ A {\bf 
223}, 1 (1996);
M.\ Lewenstein,  D.\ Bru{\ss},  J.I.\ Cirac,  B.\ Kraus,  M.\ Kus,  J.\ 
Samsonowicz,  A.\ Sanpera,  and
R.\ Tarrach, J.\ Mod.\ Opt.\ {\bf  47}, 2481 (2000);
 O.\ Rudolph, 
J.\ Phys.\ A {\bf 33}, 3951 (2000);
M.\ Horodecki, P.\ Horodecki, and R.\ Horodecki,
in {\it Quantum information: An introduction to basic theoretical concepts 
and 
experiments}, ed.\ G.\ Alber et al.\ (Springer, Heidelberg, 2001), pp. 151;
B.M.\ Terhal, Theor.\ Comput.\ Sci.\ {\bf 287}, 313 (2002);
K.\ Eckert, O.\ G{\"u}hne, F.\ Hulpke, P.\ Hyllus, J.\ Korbicz, J.\ 
Mompart, D.\ Bru{\ss}, M.\ Lewenstein, and
A.\ Sanpera, Quantum Information Processing, G.\ Leuchs and
T.\ Beth (Eds) (Wiley-VCH, Weinheim,   2003).
%
\bibitem{ConvexButNotSemidef}
K.\ Audenaert, J.\ Eisert, E.\ Jane,
M.B.\ Plenio, S.\ Virmani, and B.\ de Moor,
Phys.\ Rev.\ Lett.\ {\bf 87},  217902  (2001).
%
\bibitem{Rains}
E.M.\ Rains, IEEE Trans.\ Inf.\ Theory {\bf 47}, 2921 (2001).
%
\bibitem{ConvexInQuantumInfo}
M.\ Jezek, J.\ Rehacek, and J.\ Fiurasek, 
Phys.\ Rev.\ A {\bf 65}, 060301 (2002);
K.\ Audenaert and B.\ De Moor, Phys.\ Rev.\ A {\bf 65}, 030302 (2002);
F.\ Verstraete and H.\ Verschelde, Phys.\ Rev.\ Lett.\ {\bf 90}, 097901 
(2003);
K.\ Audenaert, M.B.\ Plenio, and J.\ Eisert, Phys.\ Rev.\ Lett.\ {\bf 
90},  027901 (2003);
Y.C.\ Eldar, M.\ Stojnic, and B.\ Hassabi, Phys.\ Rev.\ A {\bf 69}, 062318 
(2004); B.\ Synak,  K.\ Horodecki,  and M.\ Horodecki, quant-ph/0405149.
 %
 \bibitem{Global}
K.\ Audenaert, quant-ph/0402076, 
 Proceedings 
  Sixteenth International Symposium on Mathematical Theory of Networks and Systems (MTNS2004),
 Catholic University of Leuven, Belgium, 5-9 July 2004.

\bibitem{Doherty}
A.C.\ Doherty, P.A.\ Parrilo, and F.M.\ Spedalieri,
Phys.\ Rev.\ Lett.\  {\bf 88}, 187904  (2002);
A.C.\ Doherty, P.A.\ Parrilo, and F.M.\ Spedalieri,
Phys.\ Rev.\ A {\bf 69}, 022308 (2004).
%
\bibitem{Brasilians}
F.G.S.L.\ Brandao and R.O.\ Vianna, quant-ph/0405008,
quant-ph/0405063,
quant-ph/0405096.
%
\bibitem{Ioannou}
L.M.\ Ioannou, B.C.\ Travaglione, D.C.\ Cheung, and A.K.\ Ekert,
quant-ph/0403041.
%
\bibitem{Shor}
N.Z.\ Shor, Soviet Journal of Circuits and Systems Sciences {\bf 25}, 1 
(1987).
%
\bibitem{Kojima}
M.\ Kojima and L.\ Tuncel,
Mathematical Programming {\bf 89}, 79 (2000);
A.\ Takeda, K.\ Fujisawa, Y.\ Fukaya, and M.\ Kojima, 
Journal of Global Optimization {\bf 24},  237  (2002);
M.\ Kojima, S.\ Kim, and H.\ Waki,
{J.\ Ope.\ Res.\ Soc.\ Japan} {\bf 46},  2 (2003).
%
\bibitem{Lasserre}
J.B.\ Lasserre,
{SIAM J.\ Optimization} {\bf 11},  796 (2001).
%
\bibitem{Lasserre2}
D.\ Henrion and J.B.\ Lasserre, 
IEEE Control Systems Magazine {\bf 24},  72 (2004).
%
\bibitem{Par}
	P.A.\ Parrilo, {\it Structured semi-definite programs and semi-algebraic 
	geometry methods in robustness and optimization} (PhD thesis, California
	Institute of Technology, Pasadena, 2000).
%
 \bibitem{Semi}
L.\ Vandenberghe and S.\ Boyd, {\it Semidefinite programming}. 
SIAM Review {\bf 38},  49 (1996);
C.\ Helmberg, 
{\it Semidefinite programming},
European Journal of Operational Research {\bf 137}, 461  (2002).
 %
 
 
\bibitem{LagrangeDuality}
For any semi-definite program (functioning as the primal problem), 
in its most general form
being given by
 \begin{eqnarray}
 	\text{minimize} & c^T x,\\
	\text{subject to} &  	F_0 + \sum_{s=1}^t 
	x_s F_s\geq 0,\nonumber
\end{eqnarray}
one can formulate the Lagrange-dual problem, which is 
again a semi-definite problem. It is given by
	 \begin{eqnarray}
 	\text{maximize} & -\tr[F_0 Z ],\\
	\text{subject to} &  \tr[F_s Z]=c_s,\, s=1,...,t,\nonumber\\
	& Z\geq 0.\nonumber
\end{eqnarray}
The key point of Lagrange duality   is that 
any solution of the dual problem is a lower bound to the 
optimal solution of the primal problem. This is what is referred to as 
weak duality. 
Under certain conditions (in particular, if there is a solution $x$ satisfying
$F_0 + \sum_{s=1}^t x_s F_s>0$, the optimal values of the dual and the
primal problem are identical. In this case, which is rather the typical one,
one refers to strong duality. The idea of Lagrange duality is a powerful tool 
to formulate rigorous lower bounds to solutions of optimization problems. 

 
 \bibitem{NewJones}
 N.S.\ Jones and N.\ Linden,
 quant-ph/0407117.
 %
\bibitem{QubitRemark}
As pointed out before, for qubit systems 
the 
constraints  can further be simplified by merely requiring
\begin{eqnarray}
 	\tr[ \tr_{I\backslash j} [P^{(i)}]^2 ]  & = &  (\tr [P^{(i)} ])^2\\
	\tr[ \tr_{I\backslash j} [P^{(i)}]  ]  & = &  \tr [P^{(i)} ]
\end{eqnarray}	
 for $ j\in I$.

\bibitem{Rocky}
R.T.\ Rockefellar, {\it Convex analysis}
(Princeton University Press, Princeton, 1970).
\bibitem{Sharp}
This notion can be sharpened by employing the notion of 
weak membership \cite{ConvexBook}. 
This can be phrased as follows: we
denote for any convex set $S\subset \qq^m$
and any rational $\delta>0$
with $B(S,\delta)$ the set of all $x\in \qq^m$ 
for which there exists a $y\in S$
such that
\begin{equation}
	\|x - y\|_2\leq \delta,
\end{equation}
and with $B(S, -\delta)$ the set of all $x\in S$ for which 
$y\in S$ for all $y\in \qq^m$ with $\|x-y\|_2 \leq \delta$. So clearly,
$S$ is a strict subset of $B(S,\delta)$, and $B(S, -\delta)$  is a strict
subset of $S$. The weak membership problem allows for two alternatives:
given a rational element $x\in{\qq}^m$ and a rational
number $\delta>0$ either (i) assert that $x\in B(S,\delta)$, or
(ii) assert that $x\neq B(S,-\delta)$.
%
\bibitem{ConvexBook}
M.\ Gr{\"o}tschel, L.\ Lovasz, and A.\ Schrijver, {\it Geometric algorithms
and combinatorical optimization} (Springer, Heidelberg, 1988).
%
\bibitem{witness} 
        B.M.\ Terhal, Phys.\ Lett.\ A {\bf 271}, 319 (2000).
        %
\bibitem{barbieri} 
M.\ Barbieri, F.\ De Martini, G.\ Di Nepi,
P.\ Mataloni, G.M.\ D'Ariano, and C.\ Macchiavello,
Phys.\ Rev.\ Lett.\ {\bf 91}, 227901 (2003);
M.\ Bourennane, M.\ Eibl, C.\ Kurtsiefer,  S.\ Gaertner,
H.\ Weinfurter, O.\ G{\"u}hne, P.\ Hyllus, D.\ Bru{\ss},
M.\ Lewenstein, and  A.\ Sanpera, Phys.\ Rev.\ Lett.\ {\bf 92},
087902 (2004).
%
\bibitem{curtyPRL} M.\ Curty, M.\ Lewenstein, and N. L{\"u}tkenhaus, 
        Phys.\ Rev.\ Lett.\ {\bf 92}, 217903 (2004);
 	M.\ Curty, O.\ G{\"u}hne, M.\ Lewenstein
        and N.\ L{\"u}tkenhaus, quant-ph/0409047.
%
\bibitem{constructW} K.\ Chen and L.A.\ Wu, Phys.\ Rev.\ A 
	{\bf 69}, 022312 (2004);
        G.\ T{\'o}th, quant-ph/0406061.
        %
\bibitem{optimization} M.\ Lewenstein, B.\ Kraus, J.I.\ Cirac, 
        and P.\ Horodecki, Phys.\ Rev.\ A {\bf 62}, 052310 (2000).
        %
\bibitem{PPTRemark}        
         It was shown in Ref.\ \cite{bound} that
those states cannot be destilled, which is why they are referred
to as being   bound entangled. For more than two parties, there 
exist states which have a NPPT with respect to some splitting
which are nevertheless bound entangled \cite{dct}.
%
\bibitem{bound} M.\ Horodecki, P.\ Horodecki, and R.\ Horodecki,
    Phys.\ Rev.\ Lett.\ {\bf 80}, 5239 (1998).
    %
\bibitem{dct} W.\ D{\"u}r, J.I.\ Cirac, and R.\ Tarrach, 
        Phys.\ Rev.\ Lett.\ {\bf 83}, 3562 (1999).
              %
\bibitem{Full}
	G.\ Kimura,  Phys.\ Lett.\ A {\bf 314}, 339 (2003);
	M.S.\ Byrd and N.\ Khaneja, Phys.\ Rev.\ A {\bf 68}, 062322 (2003).
	%
\bibitem{gbes} P.\ Hyllus, C.\ Moura Alves, D.\ Bru{\ss}, 
        and Ch.\ Macchiavello, Phys.\ Rev.\ A {\bf 70},
	032316 (2004).
%
\bibitem{MREGS}
C.H.\ Bennett, S.\ Popescu, D.\ Rohrlich, J.A.\ Smolin, and A.V.\ 
Thapliyal,
Phys.\ Rev.\ A {\bf 63}, 012307 (2001).
%
\bibitem{MREGS2}
N.\ Linden, S.\ Popescu, B.\ Schumacher, and 
M.\ Westmoreland, quant-ph/9912039; E.F.\ Galvao, M.B.\ Plenio, and
S.\ Virmani, J.\ Phys.\ A {\bf 33}, 8809 (2000); S.\ Wu and Y.\ Zhang,
Phys.\ Rev.\ A {\bf 63}, 012308 (2001); A.\ Ac\'{\i}n, G.\ Vidal, and J.I.\ 
Cirac,
Quant.\ Inf.\ Comp.\ {\bf 3}, 55 (2003).
%
\bibitem{Linden}
H.\ Barnum and N.\ Linden, J.\ Phys.\ A {\bf 34}, 6787 (2001).
%
\bibitem{Wei}
T.-C.\ Wei and  P.M.\ Goldbart, Phys.\ Rev.\ A {\bf  68}, 042307 (2003).
%
\bibitem{Multi}
V.\ Coffman, J.\ Kundu, and W.K.\ Wootters, Phys.\ Rev.\ A {\bf 61}, 
052306 (2000);
J.\ Eisert and 
H.J.\ Briegel, Phys.\ Rev.\ A {\bf  64}, 022306  (2001); 
D.A.\ Meyer and N.R.\ Wallach, J.\ Math.\ Phys.\
{\bf 43}, 4273 (2002);
F.\ Verstraete, J.\ Dehaene, and B.\ De Moor   
Phys.\ Rev.\ A {\bf 68}, 012103 (2003);
A.J.\ Scott, Phys.\ Rev.\ A {\bf 69}, 052330 (2004);
M.\ Hein,  J.\ Eisert, and  H.J.\ Briegel, Phys.\ Rev.\ A {\bf 69}, 062311 
(2004).
%
\bibitem{Shimony}
A.\ Shimony, Ann.\ NY Acad.\ Sci.\ {\bf 755}, 675 (1995).
%
\bibitem{Gaussian}
	L.-M.\ Duan, G.\ Giedke, J.I.\ Cirac, and P.\ Zoller,
   Phys.\ Rev.\ Lett.\ {\bf 84}, 2722 (2000);
	J.I.\ Cirac, J.\ Eisert, G.\ Giedke,  M.\
    Lewenstein, M.B.\ Plenio,
    R.F.\ Werner, and M.M.\ Wolf,
    textbook in preparation (2004).
%
\bibitem{Hof}
	H.F.\ Hofmann and S.\ Takeuchi, 
	Phys.\ Rev.\ A {\bf 68}, 032103 (2003);
	G.\ T\'oth, Phys.\ Rev.\ A {\bf 69}, 052327 (2004).
%
\bibitem{Guehne}
	O.\ G{\"u}hne, Phys.\ Rev.\ Lett.\ {\bf 92}, 117903 (2004).
%
\bibitem{Linear}
R.\ Horodecki, P.\ Horodecki, and M.\ Horodecki,
Phys.\ Lett.\ A {\bf 210}, 377 (1996);
R.\ Horodecki and M.\ Horodecki, 
Phys.\ Rev.\ A {\bf 54}, 1838 (1996);
K.G.H.\ Vollbrecht and M.M.\ Wolf, 
J.\ Math.\ Phys.\ {\bf 43}, 4299 (2002);
  G.\ Adesso,  F.\ Illuminati, 
and  S.\ De Siena,
  Phys.\ Rev.\ A {\bf 68}, 062318 (2003); 
  C.\ Moura 
Alves and D.\ Jaksch, Phys.\ Rev.\ Lett.\ {\bf 93}, 110501 (2004).
 %
\bibitem{Channel}
G.G.\ Amosov,  A.S.\ Holevo, and R.F.\ Werner, 
 Problems in Information Transmission {\bf 36}, 25 (2000);
 C.\ King and  M.B.\ Ruskai, quant-ph/0401026.
 %
\bibitem{SA90}
	H.D.\ Sherali and W.P.\ Adams, SIAM Journal on Discrete 
Mathematics {\bf 3}, 	411 (1990).
%
\bibitem{gpmanual} D.\ Henrion and  J.B.\ Lasserre, ACM Transactions on 
Mathematical 
Software {\bf 29}, 165 (2003). See also the web page  
{\text www.laas.fr/$\sim$henrion/software/gloptipoly/gloptipoly.html}.
%
\bibitem{SeDuMi} J.F.\ Sturm, Optimization Methods and Software {\bf 11}, 
625 (1999); see also the documentation of the software on the
web page 
{\text fewcal.kub.nl/sturm/software/sedumi.html}.
%
%
%
\bibitem{duer} W.\ D\"ur, G.\ Vidal, and J.I.\ Cirac, Phys.\ Rev.\ A {\bf 
62},  062314 (2000).
%
\bibitem{kempe} J.\ Kempe, D.\ Bacon, D.A.\ Lidar, and K.B.\ Whaley, 
Phys.\ Rev.\ A {\bf 63}, 042307 (2001).

\bibitem{NewAndrew}
A.C.\ Doherty, P.A.\ Parrilo, and F.M.\ Spedalieri,
quant-ph/0407143.

%
%
%

\end{thebibliography}
\end{document}